\documentclass{article}
\usepackage{graphicx} 
\usepackage{bm}
\usepackage{caption}
\usepackage{amssymb}
\usepackage{xcolor, amsmath}
\usepackage[section]{placeins}
\title{Assessing the Longitudinal Impact of Environmental Chemical Mixtures on Children's Neurodevelopment: A Bayesian Approach}

\begin{document}
\author{     Wei Jia
        \\Division of Biostatistics and Bioinformatics
        \\ University of Cincinnati
        \\ {\tt jiawi@mail.uc.edu}   
            \and
				Roman A. Jandarov
        \\Division of Biostatistics and Bioinformatics
        \\ University of Cincinnati
        \\ {\tt jandarrn@ucmail.uc.edu}
}
\maketitle

\section{Abstract}
This manuscript presents a novel Bayesian varying coefficient quantile regression (BVCQR) model designed to assess the longitudinal effects of chemical exposure mixtures on children's neurodevelopment. Recognizing the complexity and high-dimensionality of environmental exposures, the proposed approach addresses critical gaps in existing research by offering a method that can manage the sparsity of data and provide interpretable results. The proposed BVCQR model estimates the effects of mixtures on neurodevelopmental outcomes at specific ages, leveraging a horseshoe prior for sparsity and utilizing a Bayesian method for uncertainty quantification. Our simulations demonstrate the model's robustness and effectiveness in handling high-dimensional data, offering significant improvements over traditional models. The model's application to the Health Outcomes and Measures of the Environment (HOME) Study further illustrates its utility in identifying significant chemical exposures affecting children's growth and development. The findings underscore the potential of BVCQR in environmental health research, providing a sophisticated tool for analyzing the longitudinal impact of complex chemical mixtures, with implications for future studies aimed at understanding and mitigating environmental risks to child health.

\section{Introduction}
Environmental chemical exposures can affect children's health such as neurological development. One important characteristic in environmental exposure research is that humans are not exposed to an isolated single chemical. As such, modeling and understanding environmental mixtures are a priority area of the National Institute of Environmental Health Science (NIEHS) \cite{carlin2013unraveling}\cite{taylor2016statistical}. Many methods have been proposed and applied to address the mixture effect of chemical exposures to children's neurodevelopment but few have analyzed longitudinal effects of mixtures. Specifically, we want to not only estimate a static effect of chemical mixture on outcome measured at a single time point, but also estimate effect of chemical mixture on outcome at a given age. \\
Numerous approaches have been proposed to estimate mixture effects. Two popular model classes are response-surface methods and exposure-index methods \cite{mcgee2023bayesian}. Response-surface approaches model complex mixtures and their relationship with outcomes non-parametricaly. One representative approach is Bayesian Kernal Machine Regression (BKMR) \cite{bobb2015bayesian} which estimates a multi-dimensional mixture-response surface with a kernel approach. On the other hand, exposure-index methods estimates mixture effect with linear exposure mixture index. The most popular method is Weighted Quantile Sum Regression (WQS) \cite{carrico2015characterization}. Both classes of methods have achieved state-of-art results. However, they are designed to solve time-invariant problems. \\
Previous models have embedded methods for modeling chemical mixtures with time-varying models in order to solve longitudinal problems. Lagged kernel machine regression \cite{liu2018lagged, liu2018modeling} combined kernel machine with lagged windows approach to estimate effects of chemical exposures measured at multiple time points on outcome measured at a single time point. Bayesian Varying coefficient kernel machine regression (BVCKMR) \cite{liu2018bayesian} combined kernel machine regression with a mixed effect model to estimate effects of exposures measured at a single time point on outcome measure at multiple time points. It estimates mixture effect for both baseline age and longitudinal trajectory.  However, kernel methods are known to perform bad when the dimensionality of data gets high\cite{bengio2005curse}. Also, the interpretation of BVCKMR results can be difficult since the interpretation mainly depends on posterior analysis. Additionally, one known limitation of BVCKMR is while modeling high dimensional chemical exposure data, sparsity of the mixture effect plays an important role in the analysis. However, methods we mentioned above do not perform any shrinkage or variable selection to handle the sparsity.\\

In this paper, we propose a Bayesian varying coefficient quantile regression (BVCQR) which can (1) characterize longitudinal effects of chemical exposure mixtures, (2) provide succinct estimations when the dimensionality of data gets high and preserve the convenience of interpretations, (3) apply a horseshoe prior to ensure the sparsity of chemical effects estimation, (4) correct uncertainty quantification using Bayesian method (5) we illustrated the robustness of BVCQR on estimating exposure effects with a simulations study and a real-life data "HOME" cohort.

\section{Method}
In this section, we will start by introducing some related previous approaches, weighted quantile sum regression (WQS), Quantile G-Computation (QGC), mixed effect model, and Bayesian varying coefficient kernel machine regression (BVCKMR). In section 2.3, we introduce our BVCWQS, and details of how we handle sparsity with horseshoe prior are given in section 2.4.

\subsection{Weighted Quantile Sum Regression and Quantile G-Computation}
For subject i, let \(y_i\) be the corresponding outcome, \(\{z_{i1}, \ldots , z_{iM}\}\) be chemical exposures, and \(x_i\) be a vector of covariates for the \(ith\) subject where \(i=1,\ldots , N\). Exposure index models such as WQS and QGC fit a linear model as:
\begin{equation}
y_i=\alpha_0 + \beta(\theta_1q_{i1} + \ldots{} + \theta_Mq_{iM}) + \bm{x}_i^T\gamma + \epsilon_i
\end{equation}

where \(q_{ip}\) represent quantile indices of \(z_{ip}\) such as \(0, 1, 2, 3\) if \(z_{ip}\) is in the 0-25 percentile, 25-50 percentile, 50-75 percentile and 75-100 percentile. \(\beta\) represents the overall association between the weighted sum and outcome. Different strategies and constrains have been applied to obtain weights and overall association for WQS and QGC. For WQS, the constrain has been set on weights is \(\sum_{m}^{M} \theta_m = 1\) and \(\theta_m>0\). Alternatively, QGC relaxes this single directional constrain and estimates coefficients for each \(q_{ij}\). Component weights are calculated by \(\theta_j=\frac{\theta_j}{\sum_{j}^{M}\theta_M}\). These methods ease the interpretation because they characterize a single parameter overall mixture effect and a set of weights to illustrate how each chemical contribute to the overall mixture effect. 

\subsection{Mixed Effect Model}
A mixed effect model is generalization of standard linear regression model that can be used to analyze and model data which generated from multiple source of variation \cite{sas1992sas, peretz2002application}. It relates one response variable (outcome) with some explanatory  variables. What sets the mixed-effects model apart is its incorporation of both fixed and random factors. Fixed effects estimate average responses across groups, akin to a standard regression model, while random effects (e.g., individual-specific effects) accommodate natural variability in responses among different entities, facilitating response estimation for each entity in the analysis. Given that measurements within the same entity are often correlated, this correlation must be considered during modeling. The nature of correlation among repeated responses can vary, leading to diverse covariance or correlation structures. The model supports various covariance structures and allows for the estimation of effects and variance parameters. The dataset's structure can range from balanced (equal observations per entity) to unbalanced (unequal observations per entity).

\subsection{Bayesian varying coefficient kernel machine regression}
Let \(y_{ij}\) be the outcome of interest where \(i=1\ldots n\) for each subject and measured at age \(j=1 \ldots J_i\). The outcome is related with M components of environmental chemicals \(z_i=(z_{1i} \ldots z_{Mi})^T\). \(x_i=(x_{1i} \ldots x_{pi})^T\) are \(p\) numbers of clinical covariates. BVCKMR is an extension of BKMR \cite{bobb2015bayesian} for analyzing longitudinal exposure data which achieved by employing random effect model. The model can be writen as:
\begin{equation}
y_{ij}=\gamma_1+\gamma_2*age_{ij}+h_1(z_{1i},\ldots,z_{Mi})+h_2(z_{1i},\ldots, z_{Mi})*age_{ij}+\bm{x_i^T}\bm{\beta}+\bm{u_{ij}^Tb_i}+\epsilon_{ij}
\end{equation} 
where \(h_q(.)\), \(q=1, 2\) are mixture effects which estimated by using kernel method. \(h_1\) represents the mixture effect at the outcome baseline, and \(h_2\) represents the mixture effect over outcome trajectory. \(\gamma_1+\gamma_2*age_{ij}\) represents a global trend line. The model can be simplified as 
\begin{equation}
    \bm{Y}_i=\bm{X}_i\bm{\beta}+\bm{W}_i\bm{h}_i+\bm{U}_i\bm{b}_i+\bm{\epsilon}_i
\end{equation}
where \(\bm{h}\) is the vector of mixture effects for each individual. \(\bm{b}\) is the vector which contains random effects of each individual. \(\bm{X}_i\)is the covariates matrix for subject \(i\).  \(\bm{U}\) and \(\bm{W}\) are matrices represent ages of subjects. \(\bm{Y}_i\) is a vector of continues outcome for subject i, where \(\bm{Y}_i=(Y_{i1}, \ldots, Y_{iJ_i})^T\) and \(\bm{\epsilon}_i\thicksim N_{J_i}(0, \sigma^2\bm{I}_{J_i})\). $\bm{h}$ can be hard to estimate by kernel function when M becomes large due to the curse of dimensionality. Also, kernel based method is know as its high computational cost.

\subsection{Bayesian varying coefficient quantile regression}
Let \(y_{ij}\) be the outcome of interest where \(i=1\ldots n\) for each subject and measured at age \(j=1 \ldots J_i\). The outcome is related with M components of environmental chemicals \(z_i=(z_{1i} \ldots z_{Mi})^T\). \(x_i=(x_{1i} \ldots x_{pi})^T\) are \(p\) numbers of clinical covariates. Our data has exposures measured at a single time point and outcome is repeatedly measured on multiple time points. Consider a random effect model from \cite{liu2018bayesian} which can be written as 
\begin{equation}
y_{ij}=\gamma_1+\gamma_2*age_{ij}+h_1(.)+h_2(.)*age_{ij}+\bm{x_i^T}\bm{\beta}+\bm{u_{ij}^Tb_i}+\epsilon_{ij}
\end{equation} 
where \(h_l(.)\), \(l=1, 2\) are mixture effects and we defined it as:
\begin{equation}
    h_l(.) = \theta_{l1}q_{1i} + \ldots + \theta_{lM}q_{Mi} + \phi_l
\end{equation}
where \(\bm{q}_i=(q_{1i} \ldots q_{Mi})^T\) is a vector of quantile transformed \(\bm{z}_i\) such as \(q_{1i} = 0, 1, 2, 3\) if \(z_{1i}\) is in the 0-25 percentile, 25-50 percentile, 50-75 percentile or 75-100 percentile, \(\bm{\theta_l} = (\theta_{l1}, \ldots, \theta_{lM})\) is estimated coefficients for chemicals, \(\phi_l\) is the variance of the mixture effect. \(h_1\) represents the mixture effect at the baseline age, and \(h_2\) represents the mixture effect over time. \(\gamma_1+\gamma_2*age_{ij}\) represents a global trend line corresponding to fixed effects.  
Following \cite{liu2018bayesian}, we can simplify our model as:
\begin{equation}
    \bm{Y}_i=\bm{X}_i\bm{\beta}+\bm{W}_i\bm{h}_i+\bm{U}_i\bm{b}_i+\bm{\epsilon}_i
\end{equation}
where \(\bm{h}\) is the vector of mixture effects for each individual. \(\bm{b}\) is the vector which contains random effects of each individual. \(\bm{X}_i\)is the covariates matrix for subject \(i\).  \(\bm{U}\) and \(\bm{W}\) are matrices represent ages of subjects. \(\bm{Y}_i\) is a vector of continues outcome for subject i, where \(\bm{Y}_i=(Y_{i1}, \ldots, Y_{iJ_i})^T\) and \(\bm{\epsilon}_i\thicksim N_{J_i}(0, \sigma^2\bm{I}_{J_i})\). The global trend line can be included by adding columns of age and intercept in to \(\bm{X}_i\).

\[
    \bm{Y} = \begin{pmatrix}
           \bm{Y}_{1} \\
           \bm{Y}_{2} \\
           \vdots \\
           \bm{Y}_{n}
         \end{pmatrix} 
         = \begin{pmatrix}
             Y_{11} \\
             \vdots \\
             Y_{1J_1} \\
             \vdots \\
             Y_{n1} \\
             \vdots \\
             Y_{nJ_n}
         \end{pmatrix}, 
    \bm{\beta} = \begin{pmatrix}
        \beta_1 \\ \vdots \\ \beta_n
    \end{pmatrix},
    \bm{X} = \begin{pmatrix}
        \bm{X}_1 \\ \vdots \\ \bm{X}_n
    \end{pmatrix}
\]
  
\[    \bm{U} = \begin{pmatrix}
        1&age_{11} & & & & &\\
        \vdots & \vdots & & & & &\\
        & & 1 & age_{21} & & &\\
        & & \vdots & \vdots & & &\\
        & & 1 & age_{2J_{1}} & & &\\
        & & & & \ddots &  & \\
        & & & &  &  1 & age_{n1}\\
        & & &  & & \vdots & \vdots & \\
        & & & &  &  1 & age_{nJ_n}\\
        
    \end{pmatrix}\]

\[\bm{b} = \begin{pmatrix}
    b_{1, 1}\\ b_{2, 1} \\ b_{1, 2} \\ b_{2, 2} \\ \vdots \\ \vdots \\ b_{1, n} \\ b_{2, n}, 
\end{pmatrix},
\bm{W} = \begin{pmatrix}
    1 & 0 & \ldots & 0 & age_{11} & 0 & \ldots & \\
    \vdots & & & & \vdots & & &\\
    1 & 0 & \ldots & 0 & age_{1J_1} & 0 & \ldots & \\
    0 & 1 & 0 & \ldots & 0 & age_{21} & 0 & \ldots & \\
    0 & \vdots & & & & \vdots & & & \\
    0 & 1 & 0 & \ldots & 0 & age_{2J_2} & 0 & \ldots \\
    0 & 0 & 1 & 0 & \ldots & 0 & age_{31} & 0 \\
    0 & 0 & \vdots & & & & \vdots & \ddots \\
    
\end{pmatrix},
\bm{h} = \begin{pmatrix}
    h_{1, 1} \\ h_{1, 2} \\ \vdots \\ h_{1, n} \\ h_{2, 1} \\ h_{2, 2} \\ \vdots \\ h_{2, n}
\end{pmatrix}
\]
The proposed model can be expressed as a hierarchical model:
\begin{equation}
   \bm{y} | \bm{h}, \bm{X}, \bm{\beta}, \sigma^2, \bm{b} \thicksim N(\bm{X\beta} + \bm{Wh} + \bm{Ub}, \sigma^2\bm{I}) 
\end{equation}
\begin{equation}
    \bm{h} | \bm{\theta}_{1}, \bm{\theta_2}, \phi_1, \phi_2 \thicksim N_{2n}\begin{pmatrix}
        \begin{pmatrix}
            \bm{q}^T\bm{\theta}_1 \\ \bm{q}^T\bm{\theta}_2
        \end{pmatrix},
        \begin{pmatrix}
            \phi_1^2\bm{I}_n & 0 \\
            0 & \phi_2^2\bm{I}_n \\
        \end{pmatrix}
    \end{pmatrix}
\end{equation}
We applied horseshoe prior for \(\bm{\theta}_{1}, \bm{\theta}_{2}\)  to handle the sparsity\cite{carvalho2009handling} and will be illustrated in the next subsection.

\begin{equation}
    \phi_1^2, \phi_2^2 \thicksim gamma^{-1}(\alpha_0, \gamma_0)
\end{equation}
\begin{equation}
    \bm{b} | \sigma^2, \bm{D} \thicksim N_{2n}(\bm{0}, \sigma^2(\bm{I}_n \bigotimes \bm{D}))
\end{equation}
where covariance matrix D follows:
\begin{equation}
    \bm{D}^{-1} \thicksim Wishart_q(v_0, C_0)
\end{equation}
\begin{equation}
    \sigma^2 \thicksim gamma^{-1}(\alpha, \gamma)
\end{equation}
\begin{equation}
    \bm{\beta} \varpropto 1
\end{equation}
\subsection{Handling Sparsity of Mixture Effects with Horseshoe Prior}
Consider the \(\bm{\theta}_1\) and \(\bm{\theta_2}\) in equation 8 which are effects of chemicals at baseline and trajectory. We assume both of   \(\bm{\theta}_1\) and \(\bm{\theta_2}\) should be sparse. The horseshoe prior is set as for each \(\theta_{lm}\) in \(\bm{\theta}_l^T=(\theta_{l1}, \ldots, \theta_{lM})\) where \(m=1 \ldots M\)
\begin{equation}
\theta_{lm}|\lambda_{lm}, \tau_l \thicksim N(0, \lambda_{lm}^2, \tau_l)
\end{equation}
\(\tau_l\) is global parameter which shrinks all \(\theta_{lm}\) toward zero; \(\lambda_{lm}\) is local parameter which allow some \(\theta_{lm}\) to escape the shrinkage. The level of shrinkage is controlled by \(\tau_l\) as large value of \(\tau_l\) will reflect little shrinkage and value close to zero will shrink all \(\theta_{lm}\) to zero.  Follow the parameterization in \cite{piironen2017hyperprior}, we set half-t priors for \(\lambda_{lm}\) and \(\tau_l\)
\begin{equation}
    \lambda_{lm} \thicksim t^{+}(df_{\lambda}, 0, 1)
\end{equation}
\begin{equation}
    \tau_l \thicksim t^{+}(df_{\tau}, 0, a_0\phi_l^2) 
\end{equation} 
where \(df_{\lambda}\) and \(df_{\tau}\) are degree of freedoms which we set to 1. The scale parameter for \(\tau_l\) are set as scaled variance of the corresponding \(h\) function and \(\lambda_{lm}\) has unit scale. \\

The BVCQR is implemented in Stan and all posterior distributions are computed by Hamiltonian-Monte Carlo algorithm which has been known as the one of most efficient algorithms for reducing correlation between sample states and therefor ensure the stability of the estimates \cite{brooks2011handbook, betancourt2017conceptual}
\section{Simulations}
We conducted a simulation study to evaluate the performance of our proposed BVCQR model for estimating \(\bm{h}_l, l=1, 2\). We simulated 36 chemical exposures based on our "HOME" cohort. We first compute the covariance matrix \(\bm{C}_{36\times36}\) of 36 exposures. Then we simulated exposure data as \(Z_i \thicksim N_{36}(\bm{0}, \bm{C})\). Then we took the quantized version of \(\bm{Z}\) which is \(\bm{q}\) . Outcome \(Y\) is generated from the model \(y_{ij}=h_1(q_{1i}, \ldots , q_{Mi})+h_2(q_{1i}, \ldots , q_{Mi})*age_{ij}+\bm{x_i^T}\bm{\beta}+\bm{u_{ij}^Tb_i}+\epsilon_{ij}\) where \(\epsilon_{ij} \thicksim N(0,1)\), \(\bm{x}_j=(x_{1j}, x_{2,i})\), and \(x_{1i} \thicksim N(0, 1)\) and \(x_{2i} \thicksim Bernoulli(0.5)\).  We simulated data on two scenarios: \\
Scenario 1: \[\bm{h}_1(q)=5 * (0.1q_1 + 0.3q_{12} + 0.4q_{24} + 0.2q_{35})\]

\(\bm{h}_2(q) = 3*(0.1q_{9} + 0.5q_{23} + 0.2q_{27} + 0.2q_{33})\)\\
Scenario 2:
\[\bm{h}_1(q)=5 * (0.1q_1 + 0.3q_{12} -0.3q_{23} + 0.4q_{24} + 0.2q_{35})\]
\(\bm{h}_2(q) = 3*(-0.2q_{6} + 0.1q_{9} + 0.5q_{23} + 0.2q_{27} + 0.2q_{33})\)\\
Scenario 1 is designed to test the robustness of BVCWQS when modeling exposures with only positive effects on the outcome, scenario 2 is designed for exposures have both positive and negative effects. \\

For each  scenario, we simulated \(n = 100\) samples. The performance of BVCQR on estimating \(\bm{h}\) was evaluated by regressing the predicted \(\hat{\bm{h}}\) on \(\bm{h}\). Intercept, slope and root mean squared error (RMSE) for \(h_1\) and \(h_2\) are presented. Close to zero intercept, and close to one slope and \(R^2\) indicate good estimations. Results for all scenarios are presented in table 1.  The results indicate a good performance of BVCWQS. For all scenarios, the intercept ranges from -0.167 to 0.003 ; the slope ranges from 0.998 to 1.023; the $\bm{R}^2$ ranges from 0.987 to 0.993; the RMSE ranges from 0.186 to 0.309. In both scenarios, intercepts are close to zero and slope and  $\bm{R}^2$ are close to one, and their RMSE keep low, for both $\bm{h}_1$ and $\bm{h}_2$ 
\begin{center}
\begin{tabular}{| cc|c|c|c|c|c |} 
 \hline
 \multicolumn{2}{|c|}{\textbf{Simulation Scenario}}& \textbf{h} & \textbf{Intercept} & \textbf{Slope} & $R^2$ & \textbf{RMSE}\\
  \hline
1   &With hs prior & 1 & -0.167& 1.023& 0.992& 0.287\\ 
     && 2 & -0.023 & 1.003& 0.992& 0.186\\
   &Without hs prior& 1& -0.240& 1.031& 0.982&0.434\\
   && 2& -0.023& 1.004& 0.988&0.236\\ 
 \hline
2   &With hs prior& 1 & -0.048& 1.010& 0.987& 0.309\\
     && 2 & -0.003& 0.998& 0.993& 0.189\\
   &Without hs prior& 1& -0.092& 1.016& 0.974&0.434\\
   && 2& -0.008& 1.001& 0.989&0.235\\
   \hline
\end{tabular}
\captionof{table}{Simulation results, regression of $\hat{\bm{h}}$ on $\bm{h}$}\label{simulation}
\end{center}

The simulation results are further plot and illustrated in Figure 1 and Figure 2, in which we plot out the exposure effects on the outcome for each specific chemical (The coefficients for chemicals inside function $\bm{h}$) for scenario 1 and scenario 2. For each scenario, estimated effects and their $95\%$ confidence intervals from BVCQR are plotted as black point , true effects are plotted as red points in figures. In both scenarios, chemicals which are related with outcome are significantly estimated by BVCQR (95\% confidence intervals do not contain zero) and all other chemicals which are not related with outcome are not significant in the model (95\% confidence intervals contain zero). 
\begin{figure}[hbt!]
\centering
\includegraphics[width=1\linewidth]{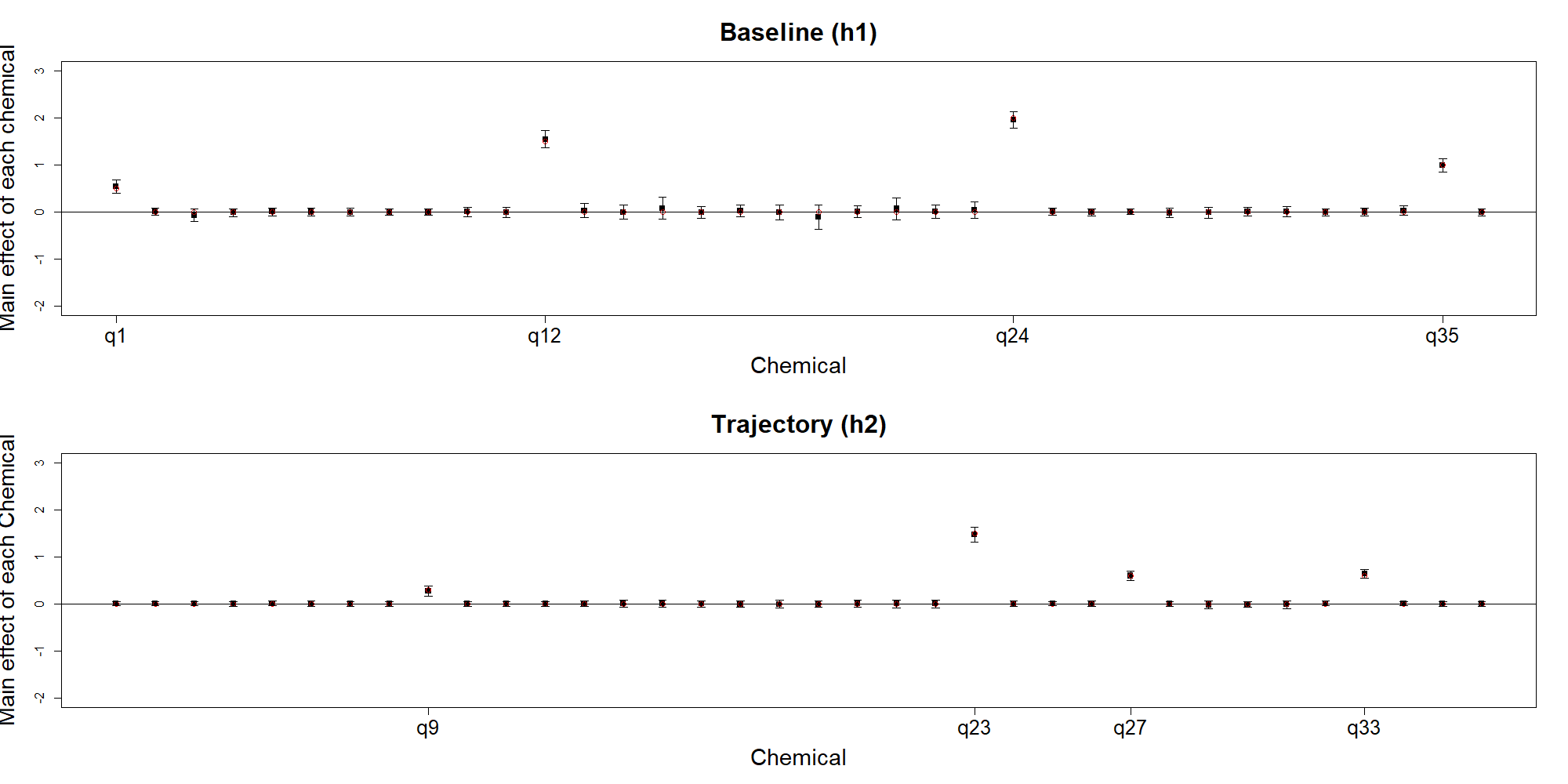}
\caption{Effects plot of Scenario 1: Upper plot is the baseline effect (h1) and the lower plot is the effect on age trajectory (h2). X axis shows names of simulated chemical exposures and Y axis shows values of each exposure effect. True effects are given as red dots and estimated effects are plot in black dots as well as their 95\% confidence. }

\end{figure}
\begin{figure}[hbt!]
\centering
\includegraphics[width=1\linewidth]{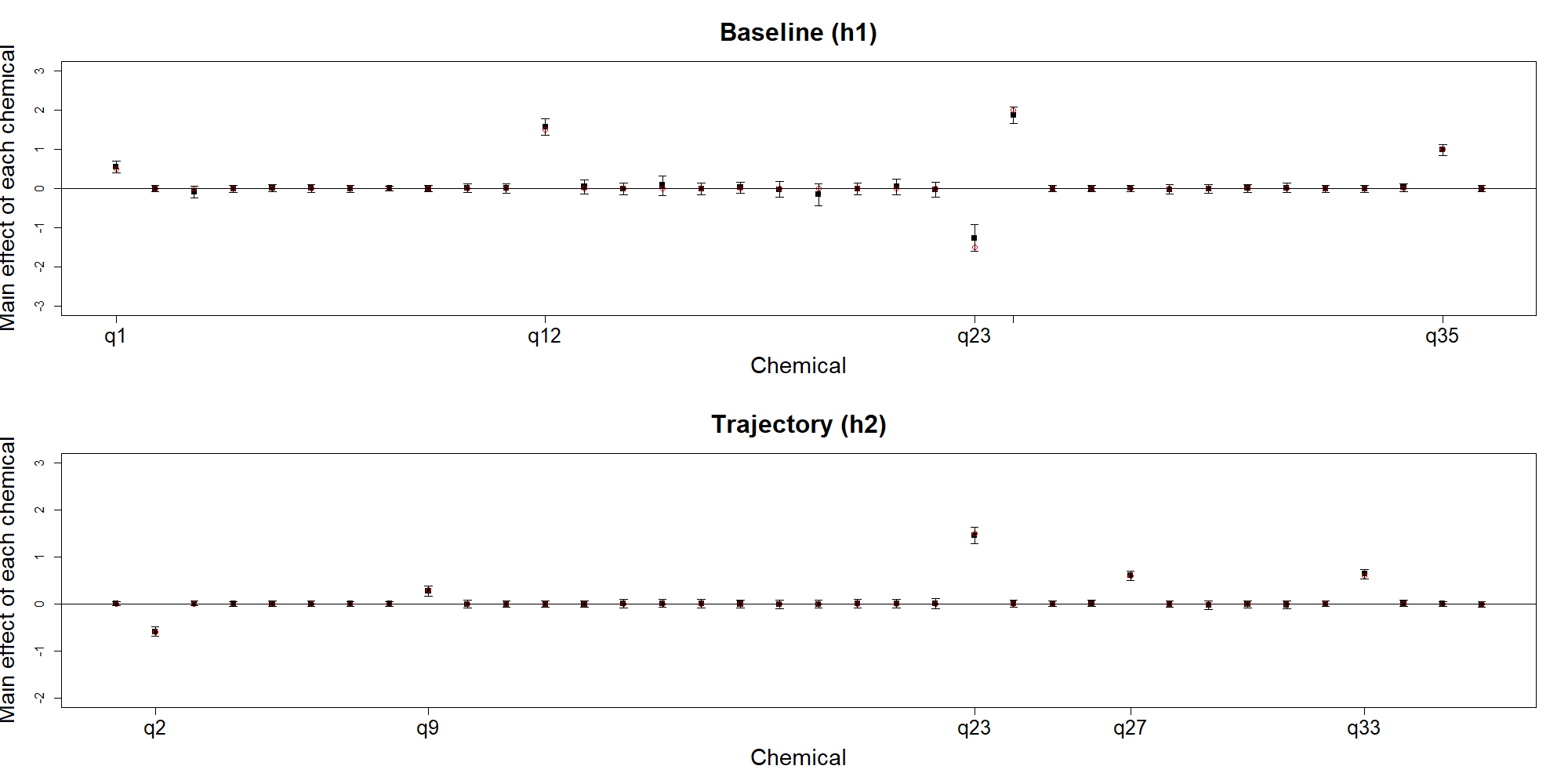}
\caption{Effects plot of Scenario 2: Upper plot is the baseline effect (h1) and the lower plot is the effect on age trajectory (h2). X axis shows names of simulated chemical exposures and Y axis shows values of each exposure effect. True effects are given as red dots and estimated effects are plot in black dots as well as their 95\% confidence.}

\end{figure}
To illustrate the sparsity handling by horseshoe prior, we conducted ablation study on simulation 1 and 2. We run BVCQR with uninformative prior specified for $\bm{\lambda}_1$ and $\bm{\lambda}_2$. The results of linear regression model of $\hat{\bm{h}}$ and $\bm{h}$ have been given in Table 1.  Our proposed BVCQR regression with horseshoe prior out performs the implementation with uninformative prior on both scenario 1 and scenario 2 on intercept, slope and $R^2$ closer to 0 and with lower RMSE. Effects estimation for scenario 1 and scenario 2 by BVCQR without horseshoe prior are given in Figure 4 and Figure 5. larger confidence intervals are estimated by BVCQR without horseshoe prior compare to BVCQR regression with horseshoe prior. Two unwanted effects, q3 and q15, which are not associated to outcome in the simulation, are estimated as significant effects at baseline (Figure 4 and Figure 5).  

\begin{figure}[hbt!]
    \centering
    \includegraphics[width=1\linewidth]{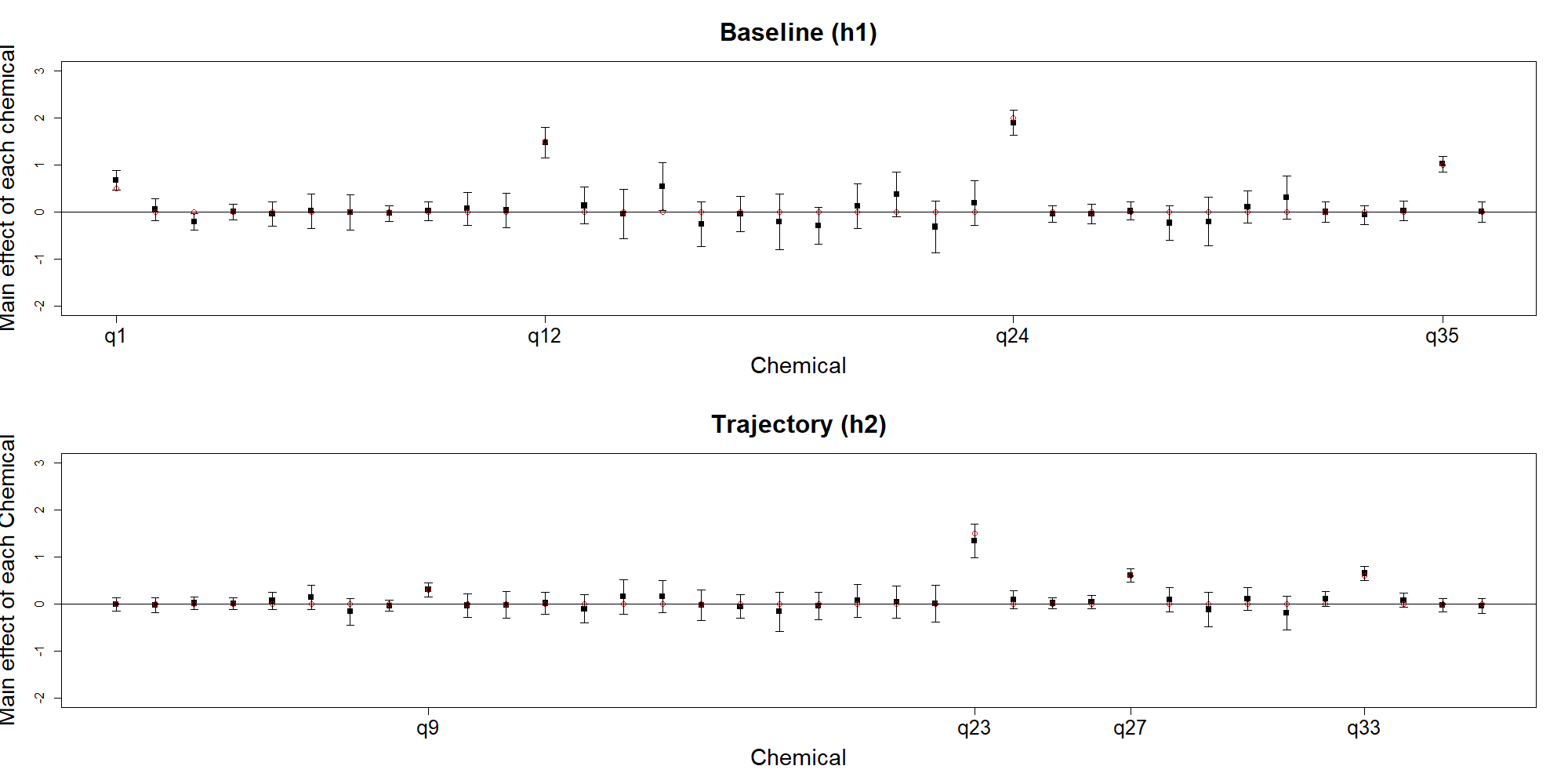}
    \caption{Effects plot of Scenario 1 estimated by BVCQR without horseshoe prior: Compare to the estimations from BVCQR with horseshoe prior, these estimations have larger confidence interval. q3 and q15 were not used to generate outcome in the simulation, however, they are falsely estimated at baseline.}
    \label{fig:enter-label}
\end{figure}
\begin{figure}[hbt!]
    \centering
    \includegraphics[width=1\linewidth]{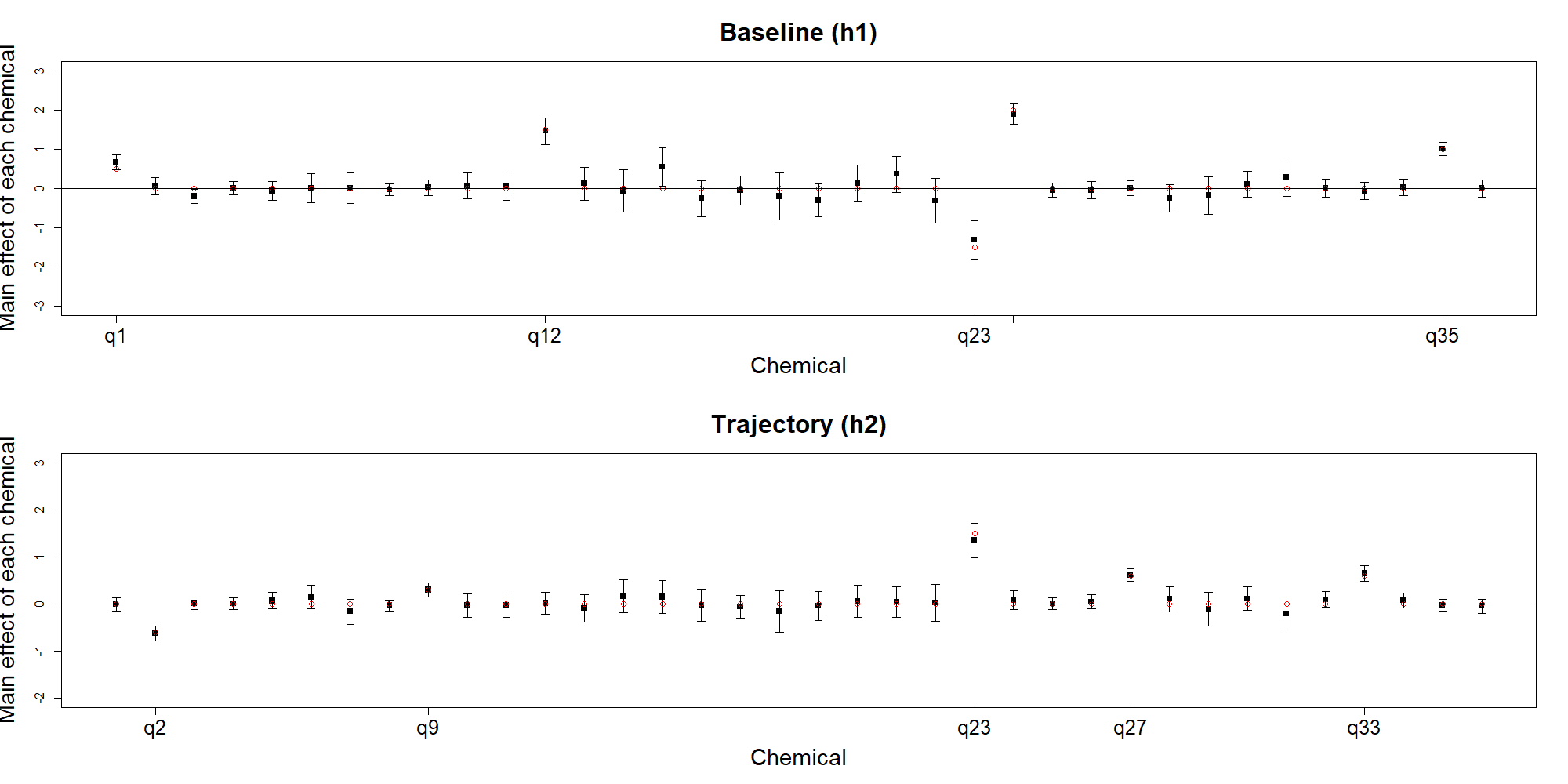}
    \caption{Effects plot of Scenario 2 estimated by BVCQR without horseshoe prior: Compare to the estimations from BVCQR with horseshoe prior, these estimations have larger confidence interval. q3 and q15 were not used to generate outcome in the simulation, however, they are falsely estimated at baseline.}
    \label{fig:enter-label}
\end{figure}
\section{Application on HOME data}
We applied BVCQR to investigate the association between chemical exposure and children's growth and development on the Health Outcomes and Measures of the environment (HOME) Study, which is a prospective cohort from three hospitals in greater Cincinnati, Ohio, metropolitan area. environmental samples were collected during pregnancy and the Bayley PDI score were collected when the children were at 12, 24 and 36 month. We used a sub-sample of this cohort (n= 147). Follow the preprocess procedure from \cite{braun2014gestational}, we excluded chemicals with less than \(20\%\) samples with detectable values and end up with 36 continuous chemicals variables. We assigned a value of \(LOD/\sqrt{2}\) for concentrations below the limit of detection\cite{hornung1990estimation}. We then rescaled these variables by dividing them by 2 times their SD\cite{gelman2008scaling}. We encoded the age 24 months as baseline. The BVCQR is implemented in Stan and compiled with R. The Hamiltonian Monte Carlo sampler was run 2000 iterations, with first 1000 used as burn-in. \\
We fitted the data with a quantile version of linear mixed effect model (LMM) with random slope and random intercept as a preliminary analysis.  We want to identify the association of chemicals and children's neurodevelopmental outcome at baseline (age 24 month) and neurodevelopmental tragectory (age 12 to 36 month) while adjustion sex, gestational age and birth weight. The baseline effects can be estimated as main exposure effects in the LMM and effects on neurodevelopmental tragectory can be estimated as the interaction effects of exposures and age. The results from LMM are given in Figure 3 to 7.  At the baseline, PFNA has positive association with Bayley PDI at 95\% significance level; at the tragectory, PCB28 has negative association at 95\% significance level.  \\ 
\begin{figure}[hbt!]
\centering
\includegraphics[width=1\linewidth]
{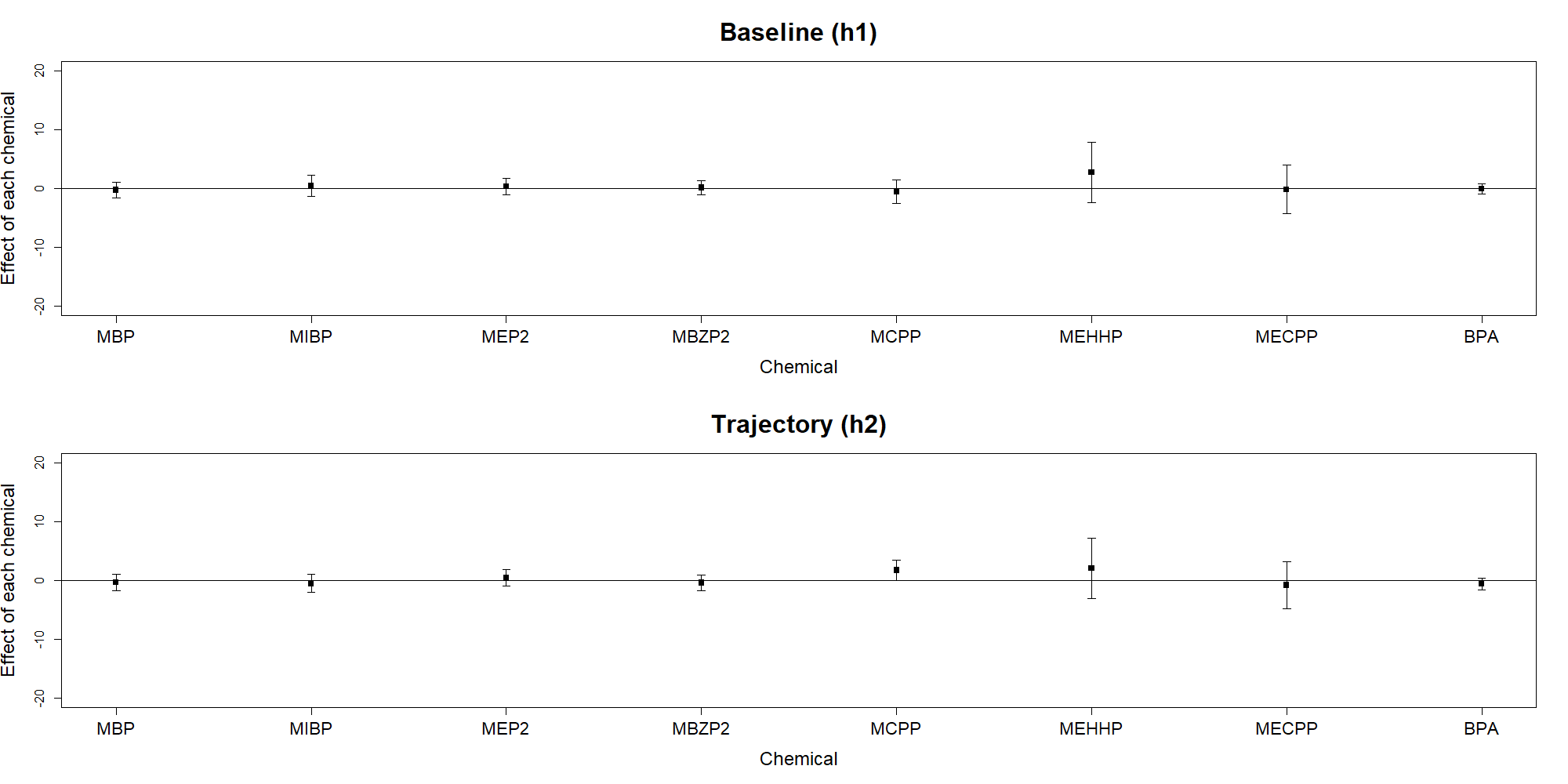}
\caption{Linear Mixed model estimated Bayley PDI at 24 months and Bayley PDI trajectory (12-36 months main effects and 95\% confidence intervals of phthbpa group chemicals.}
\end{figure}
\begin{figure}[hbt!]
\centering
\includegraphics[width=1\linewidth]
{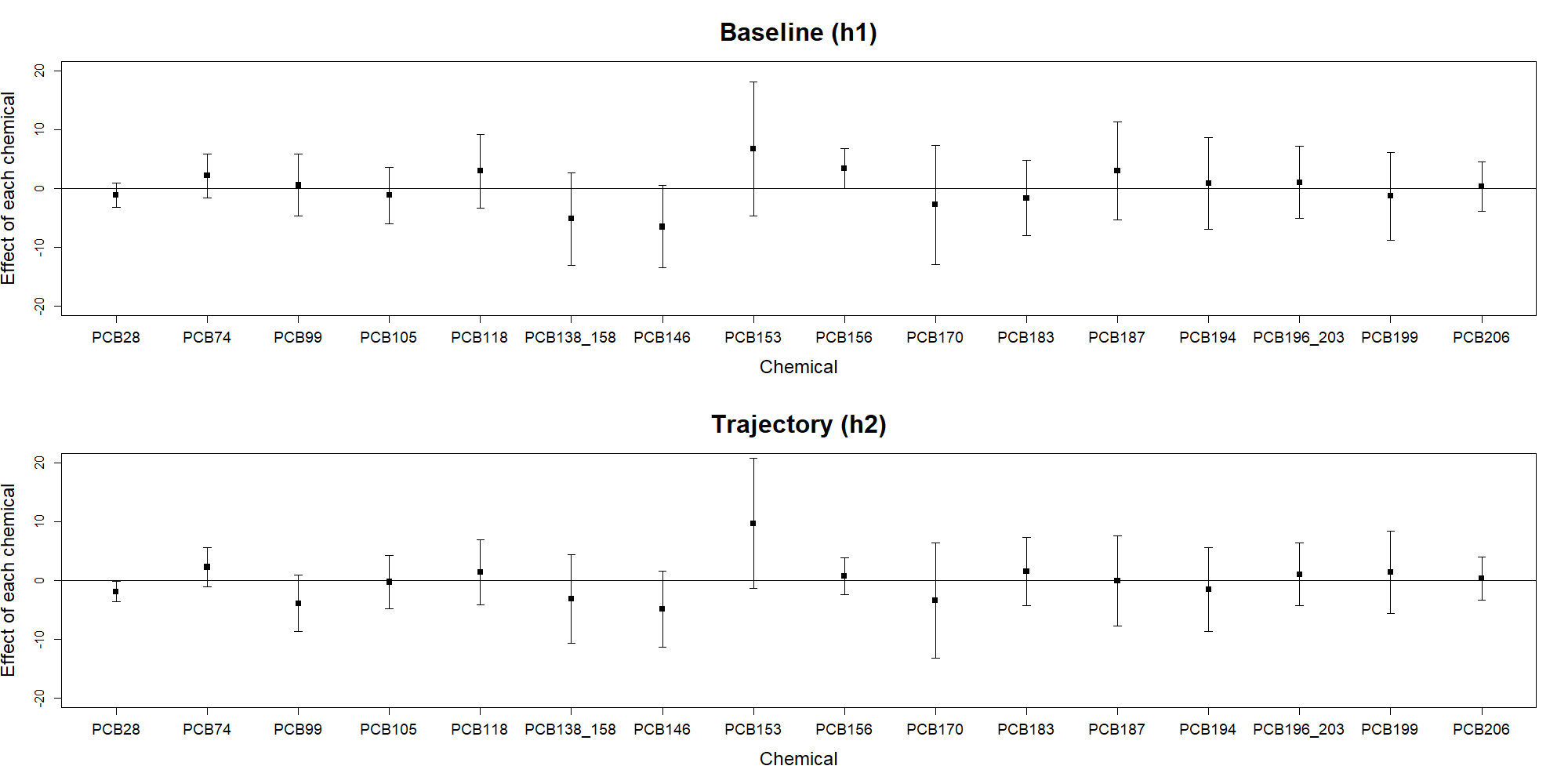}
\caption{Linear Mixed model estimated Bayley PDI at 24 months and Bayley PDI trajectory (12-36 months main effects and 95\% confidence intervals of pcb group chemicals. PCB28 has significant negative association at tragectory according to LMM estimation.}
\end{figure}
\begin{figure}[hbt!]
\centering
\includegraphics[width=1\linewidth]
{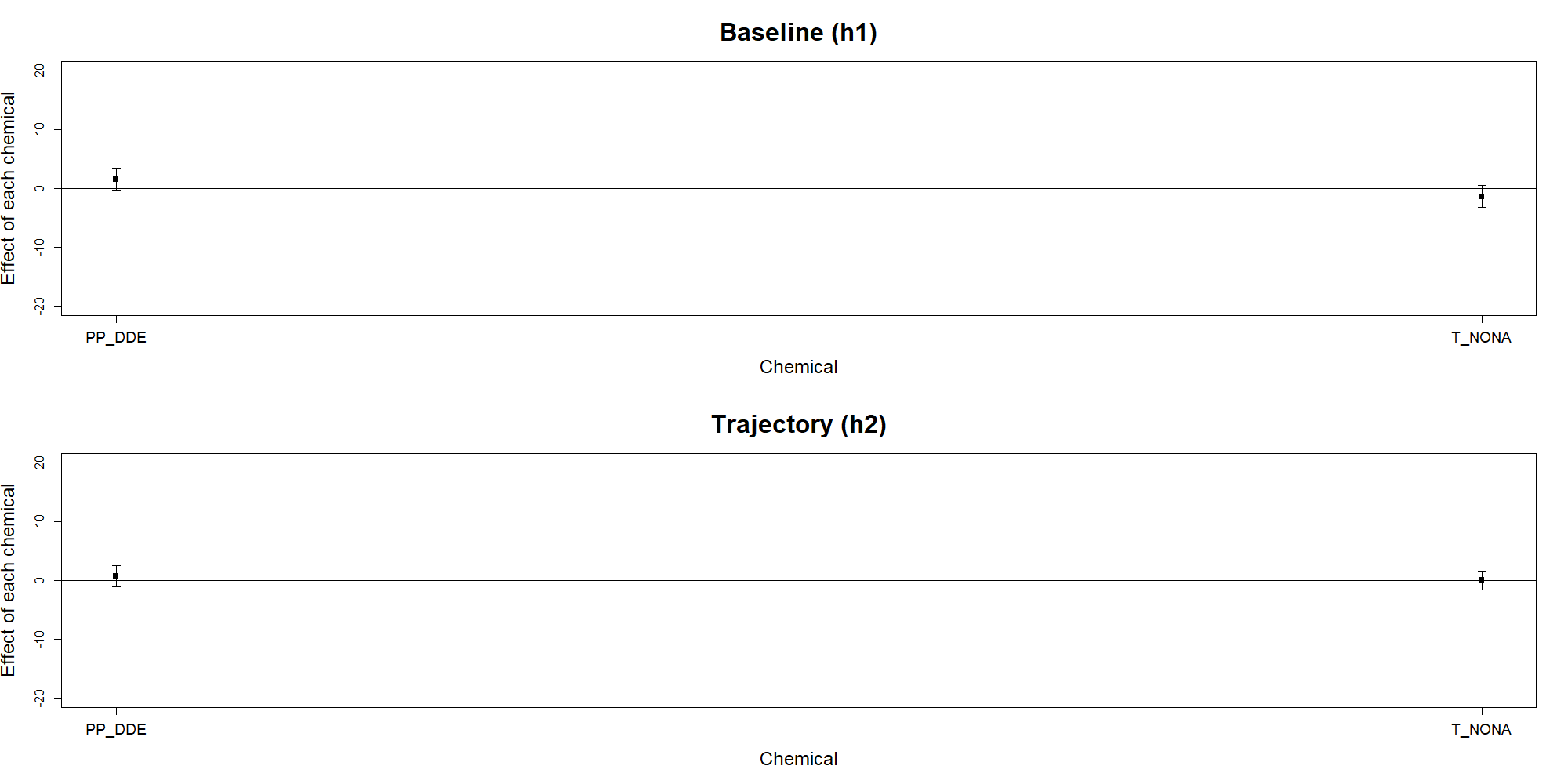}
\caption{Linear Mixed model estimated Bayley PDI at 24 months and Bayley PDI trajectory (12-36 months main effects and 95\% confidence intervals of oc group chemicals.}
\end{figure}
\begin{figure}[hbt!]
\centering
\includegraphics[width=1\linewidth]
{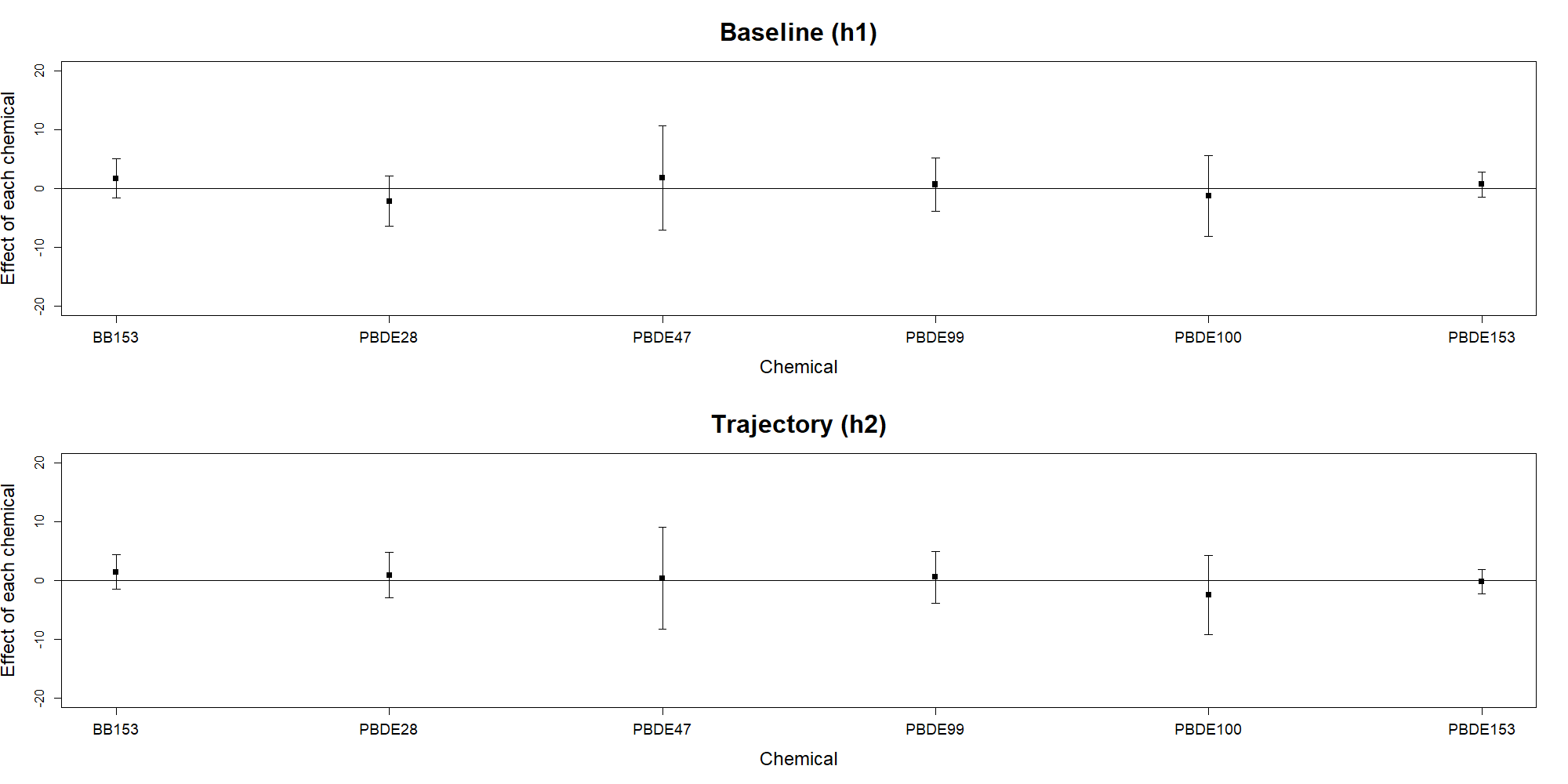}
\caption{Linear Mixed model estimated Bayley PDI at 24 months and Bayley PDI trajectory (12-36 months main effects and 95\% confidence intervals of pbde group chemicals.}
\end{figure}
\begin{figure}[hbt!]
\centering
\includegraphics[width=1\linewidth]
{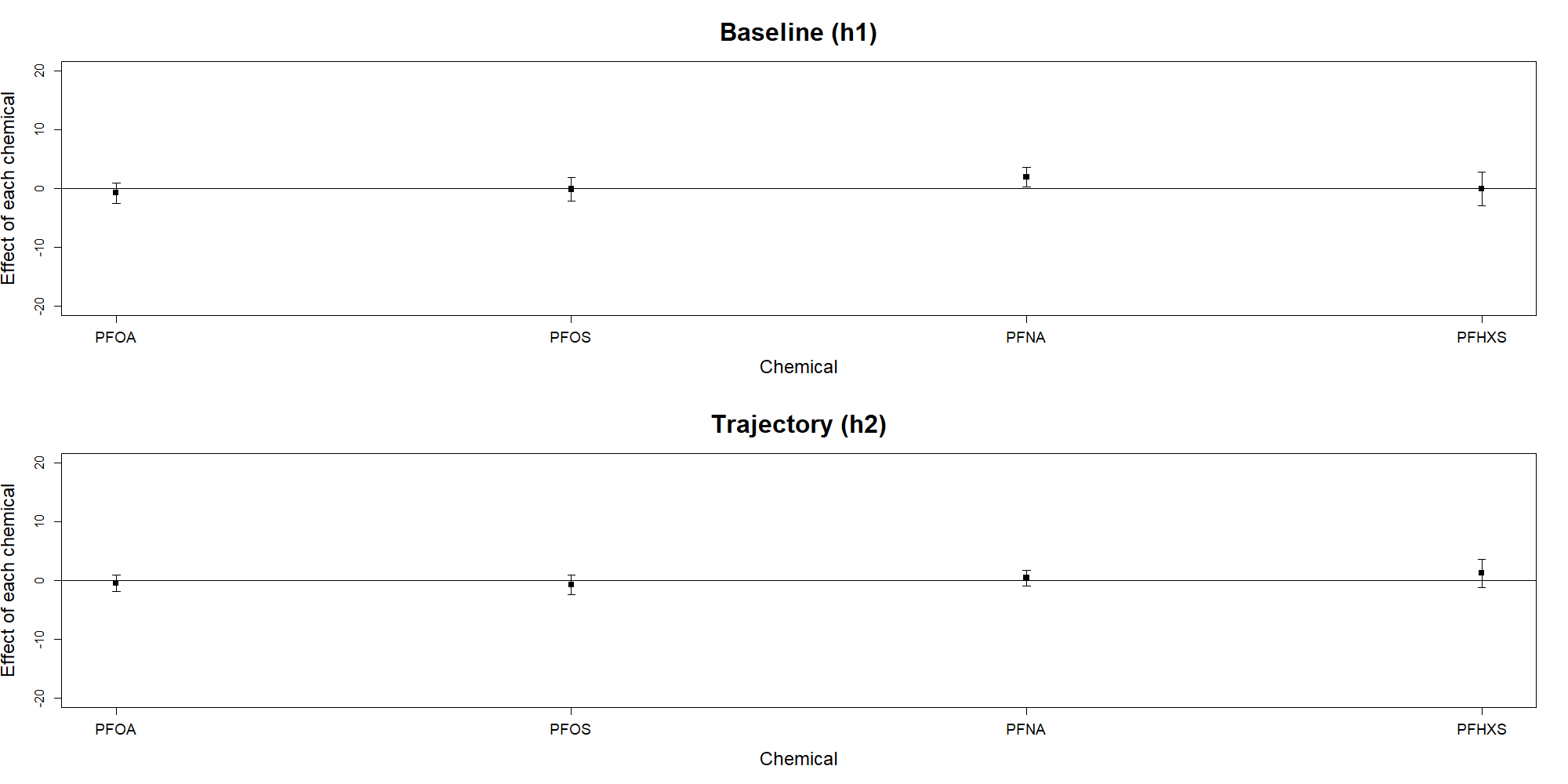}
\caption{Linear Mixed model estimated Bayley PDI at 24 months and Bayley PDI trajectory (12-36 months main effects and 95\% confidence intervals of pfc group chemicals. PFNA has significant association with Bayley PDI at 24 month.}
\end{figure}
We then use our proposed BVCQR model to fit the data. The posterior means of the parameters of the global trend line which corresponding to intercept and slope of individual-level fixed effects were $\gamma_1=1.86$ and $\gamma_2=0.33$. The effects estimations are given in Figure 10 to Figure 14. Compare to estimations from LMM, our proposed BVCQR shrinks most of insignificant chemicals to exactly zero or very close to zero. Both BVCQR and LMM detected PFNA's positive association at 24 month. Contrast to the negative association of PCB28 at tragectory which detected by LMM, BVCQR detected positive association of PCB28 at 24 month. 
\begin{figure}[hbt!]
\centering
\includegraphics[width=1\linewidth]
{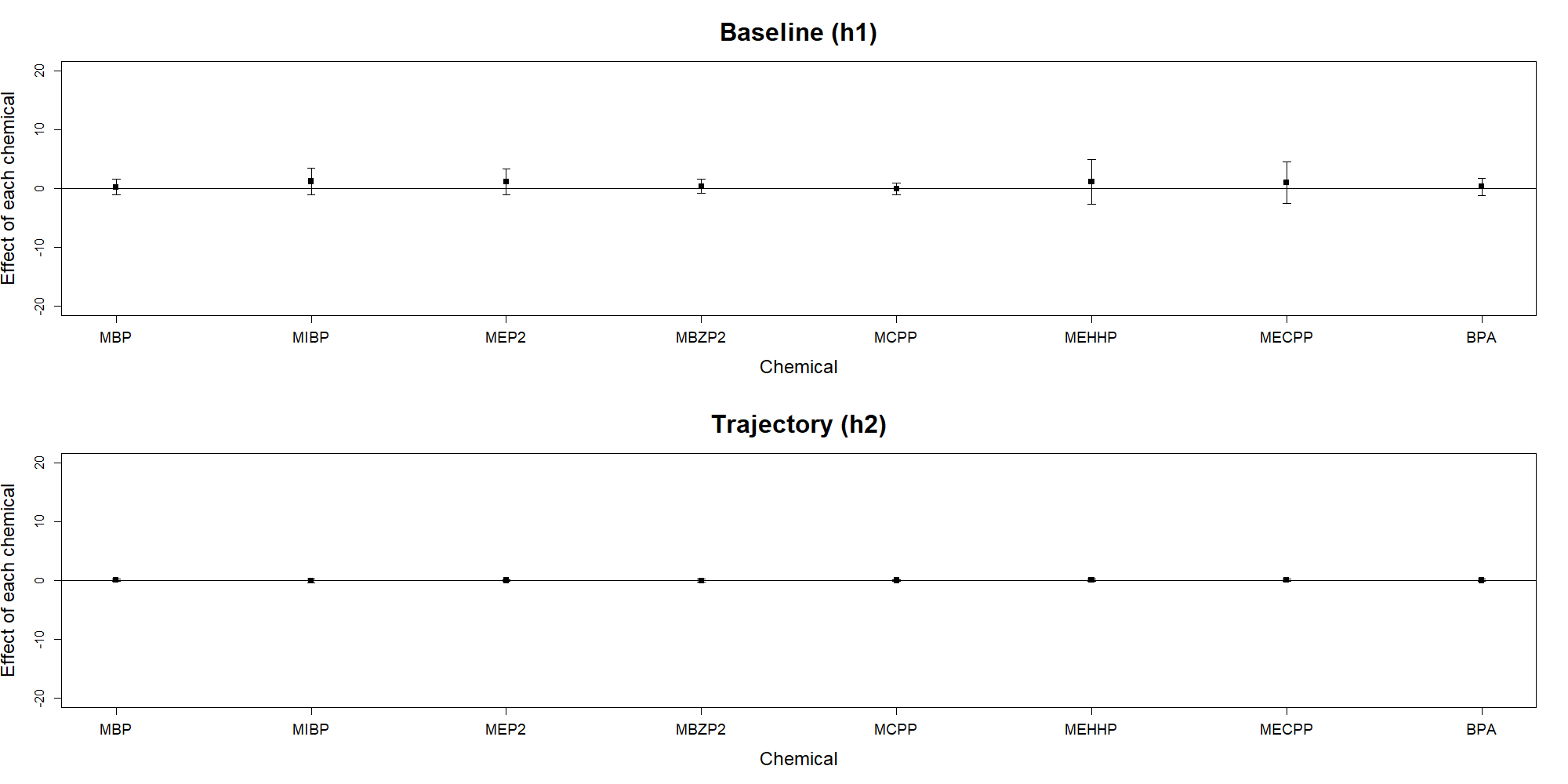}
\caption{BVCQR estimated Bayley PDI at 24 months and Bayley PDI trajectory (12-36 months main effects and 95\% confidence intervals of pfc group chemicals. phthbpa has significant association with Bayley PDI at 24 month.}
\end{figure}
\begin{figure}[hbt!]
\centering
\includegraphics[width=1\linewidth]
{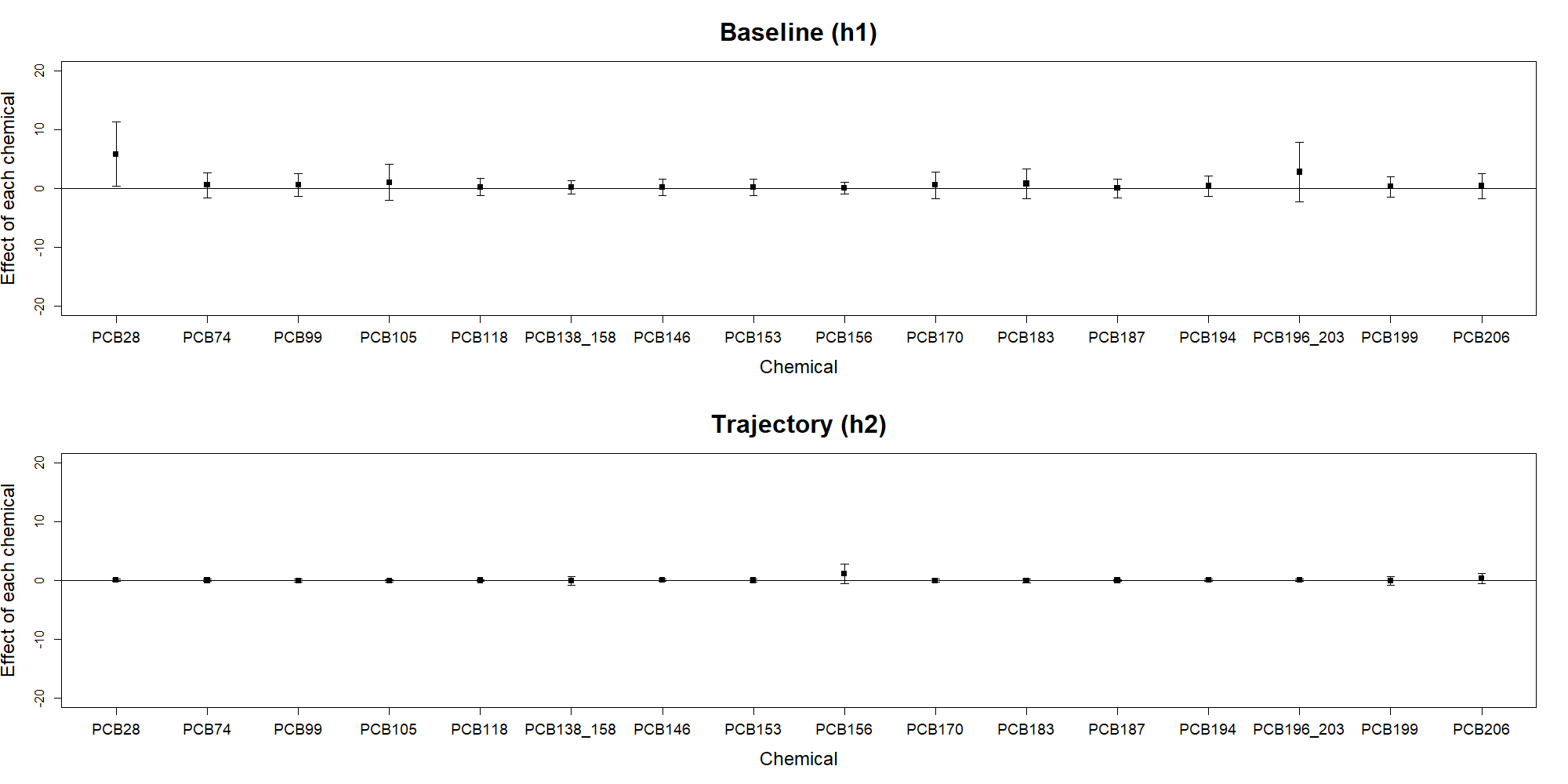}
\caption{BVCQR estimated Bayley PDI at 24 months and Bayley PDI trajectory (12-36 months main effects and 95\% confidence intervals of pcb group chemicals. PCB28 has significant positive association at tragectory according to LMM estimation.}
\end{figure}
\begin{figure}[hbt!]
\centering
\includegraphics[width=1\linewidth]
{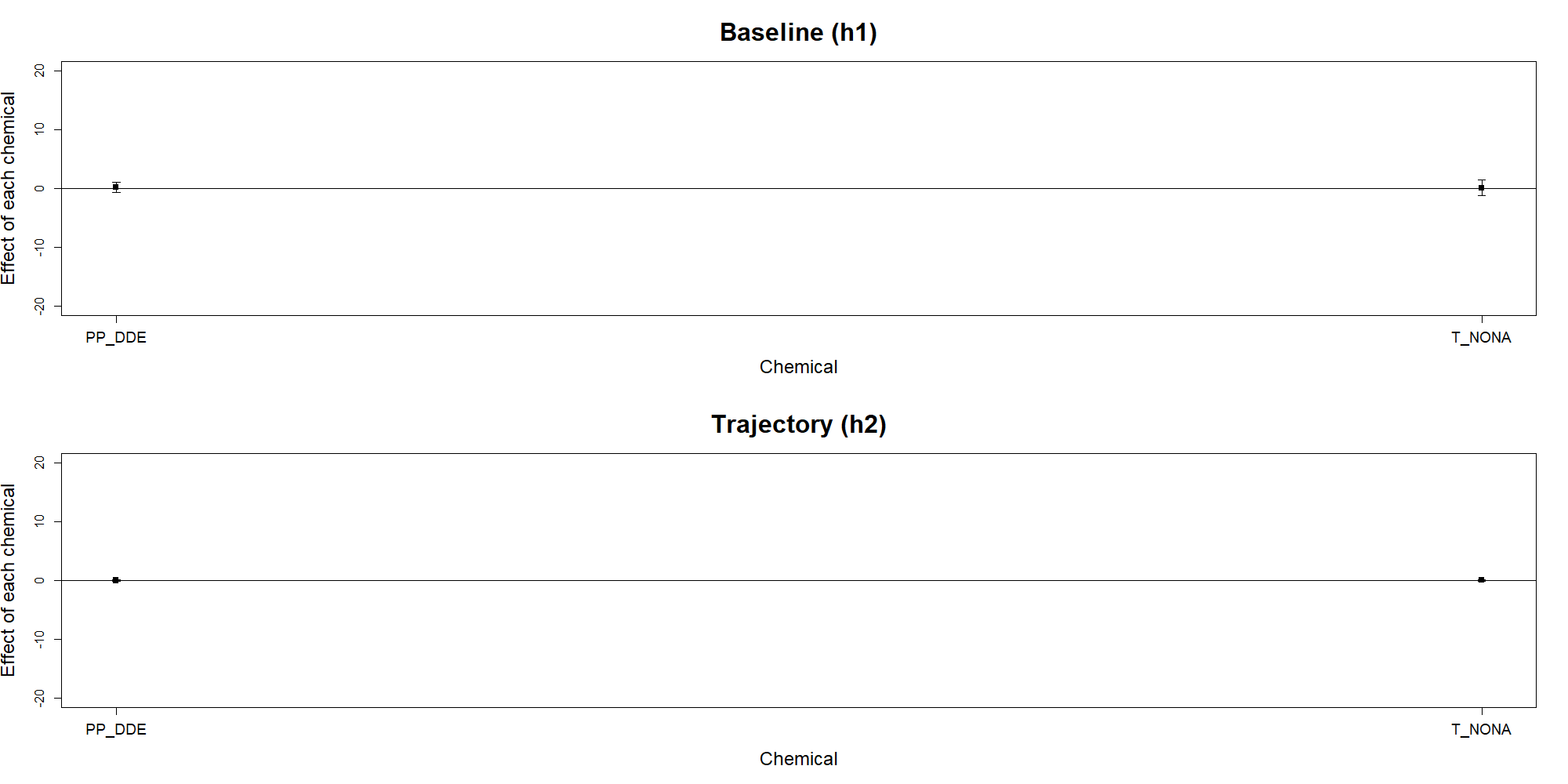}
\caption{BVCQR estimated Bayley PDI at 24 months and Bayley PDI trajectory (12-36 months main effects and 95\% confidence intervals of oc group chemicals.}
\end{figure}
\begin{figure}[hbt!]
\centering
\includegraphics[width=1\linewidth]
{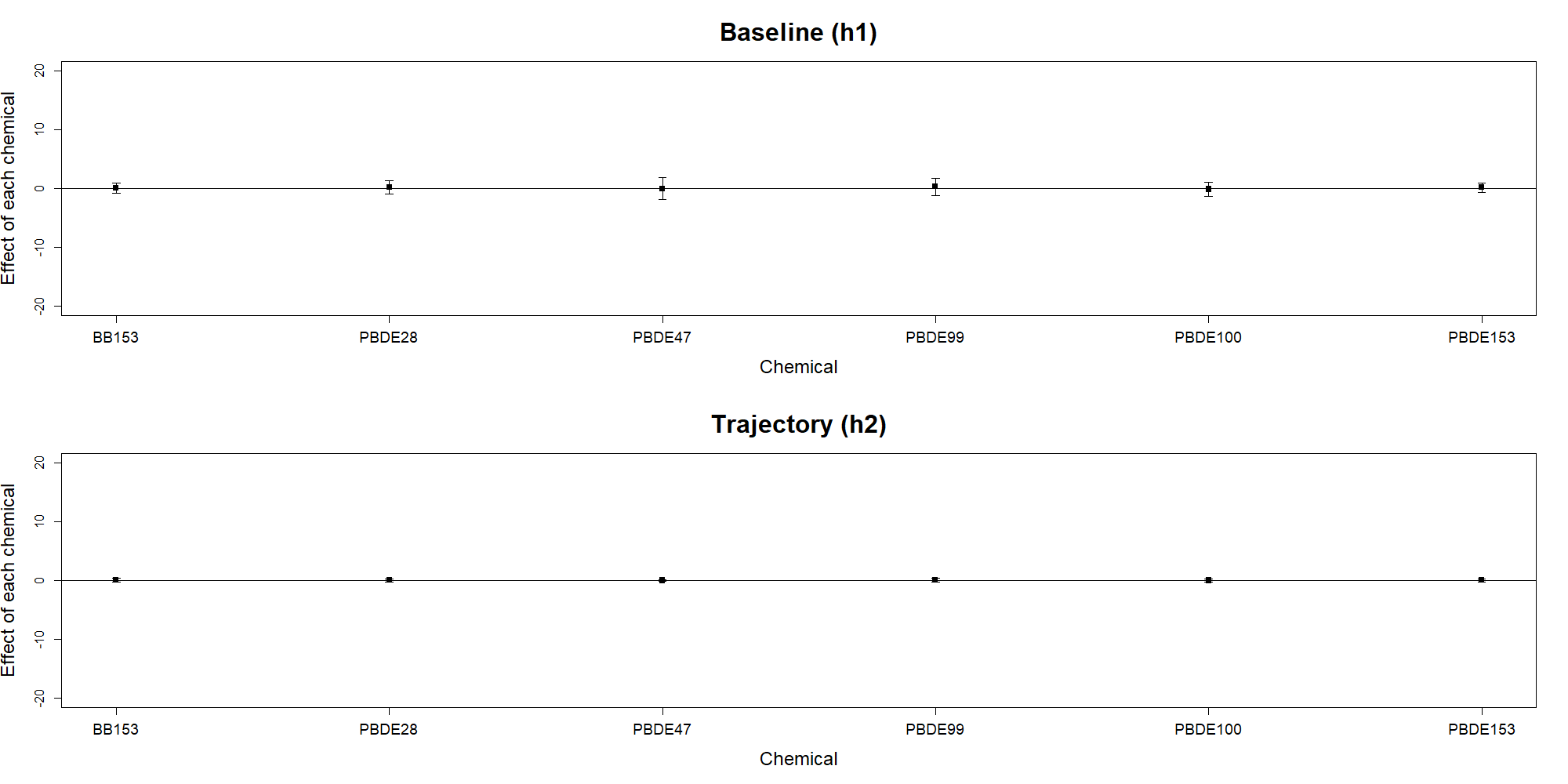}
\caption{BVCQR estimated Bayley PDI at 24 months and Bayley PDI trajectory (12-36 months main effects and 95\% confidence intervals of pbde group chemicals.}
\end{figure}
\begin{figure}[hbt!]
\centering
\includegraphics[width=1\linewidth]
{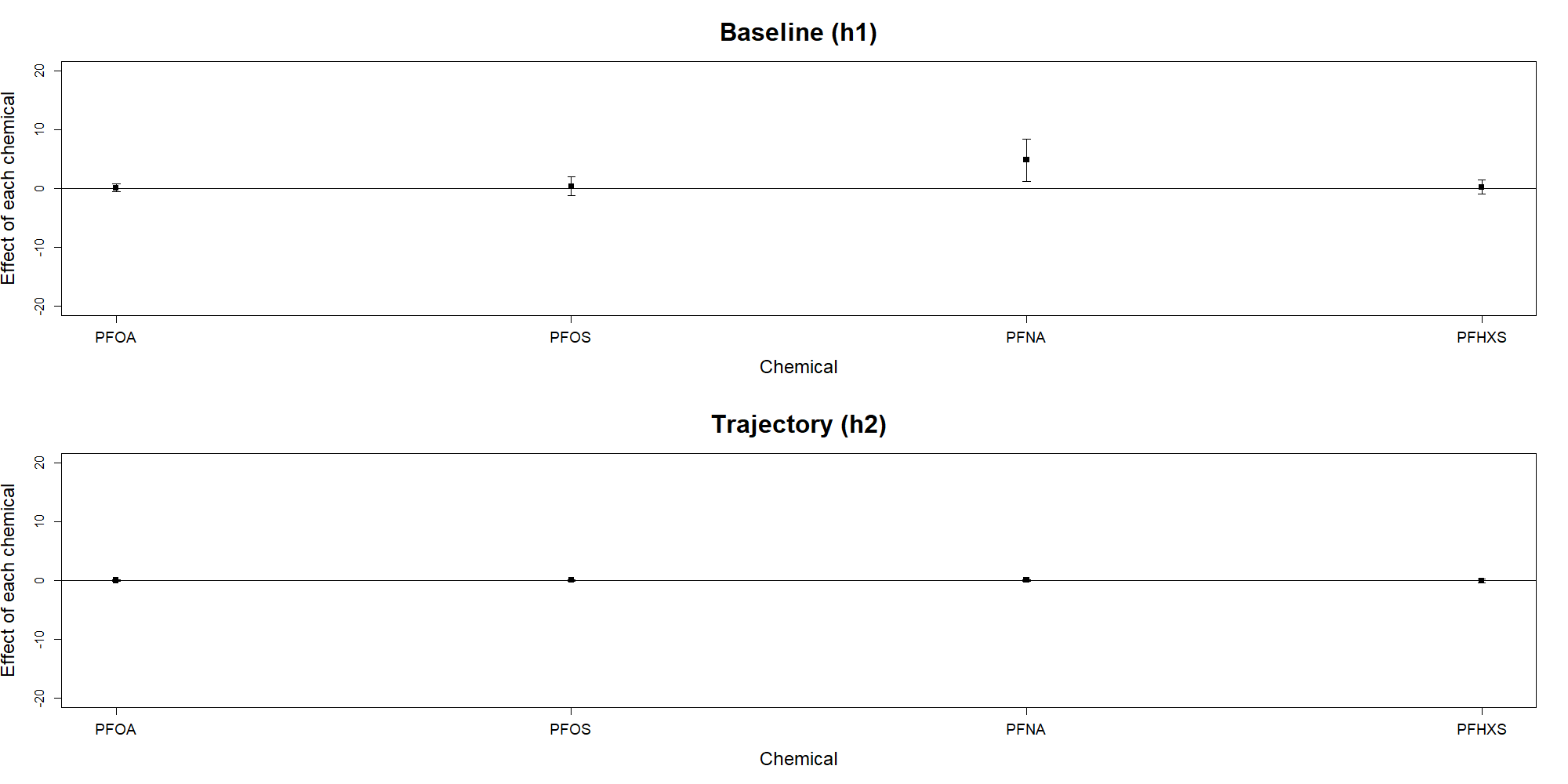}
\caption{BVCQR estimated Bayley PDI at 24 months and Bayley PDI trajectory (12-36 months main effects and 95\% confidence intervals of pfc group chemicals. PFNA has significant association with Bayley PDI at 24 month.}
\end{figure}
\section{Discussion}
We developed a Bayesian hierarchical model to analyze the effect of high dimensional chemical exposures on the outcome longitudinal trajectories. We handling sparsity of chemical exposures in $h$ function with horseshoe prior. We showed that our proposed BVCQR robustly model chemical exposure effects as well as handling sparsity on both outcome baseline and trajectory with simulations. We also included uncertainty quantification and preserved interpretbility of BVCQR. Using BVCQR to analyzing home data, we found similar set of significant chemicals compare to a quantile version linear mixed effect model, while our BVCQR shrinks effects of all insignificant chemicals to either zero or very close to zero.\\
There are also some other factors which can be considered and studied in the future to improve the function of BVCQR: (1) chemicals are belong to different groups, chemicals belong to the same group are with similar chemical structures which can lead to similar effect on the health outcomes. This information can be used to perform the variable selection. (2) interaction and non-linear effect can be estimated by incorporating other strategies. Previously, people have applied Multiple Index model \cite{mcgee2023bayesian} on exposure data to deal with these two factors. We are exploring to incorporate multiple index model with BVCQR to solve these two problems under the longitudinal environment. 

\bibliographystyle{plain}
\bibliography{ref}
\end{document}